\documentclass[reprint,showpacs,preprintnumbers,amsmath,amssymb,twocolumn,superscriptaddress]{revtex4}

	\usepackage{graphicx,dcolumn,bm,upgreek}

\begin{document}
\title{Spatial Fluctuations Of Fluid Velocities \\In Flow Through A Three-Dimensional Porous Medium}

    \author{Sujit S. Datta}
    \affiliation{Department of Physics, Harvard University, Cambridge MA 02138, USA}           

    \author{H. Chiang}
            \affiliation{Department of Physics, Harvard University, Cambridge MA 02138, USA}

    \author{T. S. Ramakrishnan}
    \affiliation{Schlumberger-Doll Research, Cambridge, MA 02138, USA}

    \author{David A. Weitz}
    \affiliation{Department of Physics, Harvard University, Cambridge MA 02138, USA}
\email{weitz@seas.harvard.edu}

\date{\today}

\begin{abstract}   
We use confocal microscopy to directly visualize the spatial fluctuations in fluid flow through a three-dimensional porous medium. We find that the velocity magnitudes and the velocity components both along and transverse to the imposed flow direction are exponentially distributed, even with residual trapping of a second immiscible fluid. Moreover, we find pore-scale correlations in the flow that are determined by the geometry of the medium. Our results suggest that, despite the considerable complexity of the pore space, fluid flow through it is not completely random.
 
\end{abstract}
\maketitle

Filtering water, squeezing a wet sponge, and brewing coffee are all familiar examples of forcing a fluid through a porous medium. This process is also crucial to many technological applications, including oil recovery, groundwater remediation, geological CO$_{2}$ storage, packed bed reactors, chromatography, fuel cells, chemical release from colloidal capsules, and even nutrient transport through mammalian tissues \cite{bear,yang,giddings,fuel,datta,tissue1,ranft}. Such flows, when sufficiently slow, are typically modeled using Darcy's law, $|\Delta p|=\mu qL/k$, where $\mu$ is the fluid dynamic viscosity and $k$ is the absolute permeability of the porous medium; this law relates the pressure drop $\Delta p$ across a length $L$ of the entire medium to the flow velocity $q$, averaged over a sufficiently large length scale. However, while appealing, this simple continuum approach neglects local pore-scale variations in the flow, which may arise as the fluid navigates the tortuous three-dimensional (3D) pore space of the medium. Such flow variations can have important practical consequences; for example, they may result in spatially heterogeneous solute transport through a porous medium. This impacts diverse situations ranging from the drying of building materials \cite{efflorescence}, to biological flows \cite{tissue1,tissue2,ranft}, to geological tracer monitoring \cite{bear}. Understanding the physical origin of these variations, on scales ranging from that of an individual pore to the scale of the entire medium, is therefore both intriguing and important. 

Experimental measurements using optical techniques \cite{Saleh1,Saleh2,Cenedese,Pinder,Rashidi,Moroni,Moroni2,Lachhab,Huang,Hassan,Arthur,Mitra} and nuclear magnetic resonance imaging \cite{Shattuck,Kutsovsky,Lebon1,Lebon2,sederman1,sederman2} confirm that the fluid speeds are broadly distributed. However, these measurements often provide access to only one component of the velocity field, and only for the case of single-phase flow; moreover, they typically yield limited statistics, due to the difficulty of probing the flow in 3D, both at pore scale resolution and over large length scales. While theoretical models and numerical simulations provide crucial additional insight \cite{Lebon1,Lebon2,Noble,Genabeek,Maier1,Maier,Talon,Zaman,Martys,auzerais,bear,scheidegger,dejong,saffman}, fully describing the disordered structure of the medium can be challenging. Consequently, despite its enormous practical importance, a complete understanding of flow within a 3D porous medium remains elusive. 

In this Letter, we use confocal microscopy to directly visualize the highly variable flow within a 3D porous medium over a broad range of length scales, from the scale of individual pores to the scale of the entire medium. We find that the velocity magnitudes and the velocity components both along and transverse to the imposed flow direction are exponentially distributed, even when a second immiscible fluid is trapped within the medium. Moreover, we find underlying pore-scale correlations in the flow, and show that these correlations are determined by the geometry of the medium. The pore space is highly disordered and complex; nevertheless, our results indicate that fluid flow through it is not completely random.

We prepare a rigid 3D porous medium by lightly sintering a dense, disordered packing of hydrophilic glass beads, with radii $32\pm2~\upmu$m, in a thin-walled square quartz capillary of cross-sectional area $A=9$ mm$^{2}$. The packing has length $L\approx8$ mm and porosity $\phi\approx0.41$, as measured using confocal microscopy; this corresponds to a random loose packing of frictional particles. Scattering of light from the surfaces of the beads typically precludes direct observation of the flow within the medium. We overcome this limitation by formulating a mixture of $82~\mbox{wt}$\% glycerol, $12~\mbox{wt}$\% dimethyl sulfoxide, and $6~\mbox{wt}$\% water, laden with 0.01 vol\% of 1 $\upmu$m diameter fluorescent latex microparticles; this composition matches the fluid refractive index with that of the glass beads, enabling full visualization of the flow through the pore space \cite{amber}. Prior to each experiment, the porous medium is evacuated under vacuum and saturated with CO$_{2}$ gas, which is soluble in the tracer-laden fluid; this procedure eliminates the formation of trapped bubbles. We then saturate the pore space with the tracer-laden fluid, imposing a constant volumetric flow rate $Q=0.2$ mL/hr; the average interstitial velocity is given by $q/\phi\equiv(Q/A)/\phi=15~\upmu$m/s, and thus the typical Reynolds number is $<10^{-3}$. The tracer Peclet number, quantifying the importance of advection relative to diffusion in determining the particle motion, is $>10^{3}$. 

 \begin{figure}
\begin{center}
\includegraphics[width=3.4in]{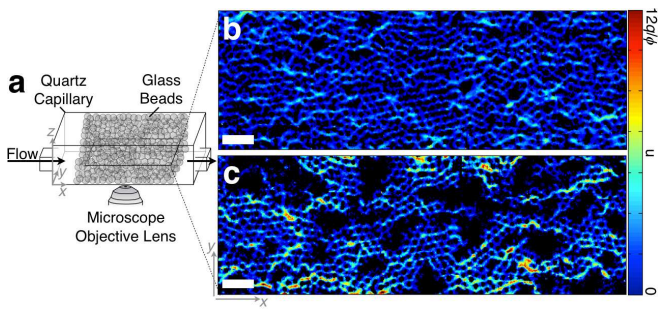}
\caption{(a) Schematic showing porous medium and imaging geometry. A portion of the 2D map of velocity magnitudes within the medium is shown for (b) single-phase flow, and (c) flow with trapped residual oil. Black circles in (b) show beads making up the medium, while additional black regions in (c) show residual oil. Scale bars are 500 $\upmu$m long, while color scale shows speeds ranging from 0 (blue) to 12$q/\phi$ (red).} 
\end{center}
\end{figure}

To directly visualize the steady-state pore-scale flow \cite{supp}, we use a confocal microscope to acquire a movie of 100 optical slices in the $xy$ plane, collecting 15 slices/s, at a fixed $z$ position several bead diameters deep within the porous medium. Each slice is 11 $\upmu$m thick along the $z$ axis and spans a lateral area of 912 $\upmu$m $\times$ 912 $\upmu$m in the $xy$ plane [Figure 1(a)]. To visualize the flow at the scale of the entire medium, we acquire additional movies, at the same $z$ position, but at multiple locations in the $xy$ plane spanning the entire width and length of the medium. We characterize the flow field using particle image velocimetry, dividing each optical slice into 16129 interrogation windows, and calculating the displacement of tracer particles in each window by cross-correlating successive slices of each movie. By combining the displacement field thus obtained for all the positions imaged, and dividing the displacements by the fixed time difference between slices, we generate a map of the two-dimensional (2D) fluid velocities, ${\bf u}$, over the entire extent of the porous medium. This protocol thus enables us to directly visualize the flow field, both at the scale of the individual pores and at the scale of the overall medium. 

 \begin{figure}
\begin{center}
\includegraphics[width=3.96in]{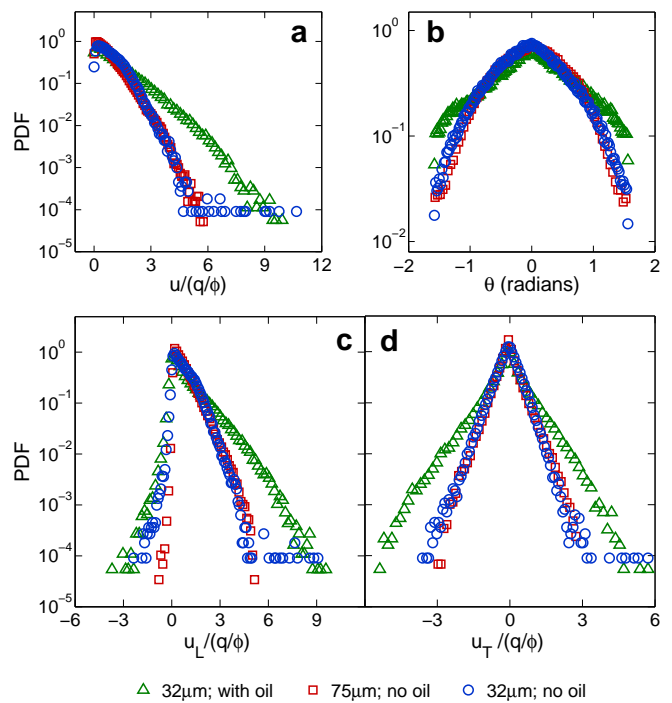}
\caption{Probability density functions of 2D (a) normalized velocity magnitudes, $u/(q/\phi)$, (b) velocity orientation angles, $\theta$, (c) normalized velocity component along the imposed flow direction, $u_{L}/(q/\phi)$, and (d) normalized velocity component transverse to the imposed flow direction, $u_{T}/(q/\phi)$. Blue circles and red squares show statistics of single-phase flow through media comprised of $32~\upmu$m and $75~\upmu$m radius beads, respectively, while green triangles show statistics of flow through the medium comprised of $32~\upmu$m radius beads, with trapped residual oil. }  
\end{center}
\end{figure}

The flow within the porous medium is highly variable, as illustrated by the map of velocity magnitudes shown in Fig.~1(b). To quantify this behavior, we calculate the probability density functions (pdfs) of the 2D velocity magnitudes, $u=|{\bf u}|$, velocity orientation angles relative to the imposed flow direction, $\theta$, and the velocity components both along and transverse to the imposed flow direction, $u_{L}=u\cos\theta$ and $u_{T}=u\sin\theta$, respectively. Consistent with the variability apparent in Fig.~1(b), we find that both the velocity magnitudes and orientations are broadly distributed, as shown by the blue circles in Fig.~2(a-b). Interestingly, the pdf of $u$ decays nearly exponentially, with a characteristic speed $\approx0.5q/\phi$, consistent with the results of recent numerical simulations \cite{branko}.

The pore space is highly disordered and complex; as a result, we expect flow through it to be random, and thus, the motion of the fluid transverse to the imposed flow direction to be Gaussian distributed \cite{saffman,fleurant}. As expected, the measured pdf of $u_{T}$ is symmetric about $u_{T}=0$; however, we find that it is strikingly non-Gaussian, again exhibiting an exponential decay over nearly four decades in probability, with a characteristic speed $\approx0.25q/\phi$ [blue circles in Fig.~2(d)]. The pdf of $u_{L}$ similarly decays exponentially, consistent with results from previous NMR measurements \cite{Kutsovsky,tail}; moreover, the characteristic speed along the imposed flow direction is $\approx 0.5q/\phi$, double the characteristic speed in the transverse direction [blue circles in Fig.~2(c)]. These results indicate that flow within a 3D porous medium may, remarkably, not be completely random.

To elucidate this behavior, we characterize the spatial structure of the flow by examining the length scale dependence of the statistics shown in Fig.~2. We do this by calculating the velocity pdfs for observation windows, centered on the same pore, of different sizes. Similar to the pdfs for the entire medium, the pdfs for windows one pore in size are broad; however, they have a different shape, as exemplified by the diamonds in Fig.~S1 \cite{supp}. By contrast, the pdfs for larger observation windows, even those just a few pores in size, are similar to those for the entire medium; two examples are shown by the crosses and stars in Fig.~S1, corresponding to windows two and ten pores across, respectively. This suggests that the variability of flow within the entire porous medium reflects a combination of the flow variability within the individual pores and the geometry of the pore space.

Another clue to the physical origin of this non-random behavior comes from close inspection of the flow field in Fig.~1(b): we observe tortuous ``fingers", approximately one pore wide and extending several pores along the imposed flow direction, over which the velocity magnitudes appear to be correlated. To quantify these correlations, we subtract the mean velocity from each 2D velocity vector to focus on the velocity fluctuations, $\delta{\bf u}$; we then calculate a spatial correlation function that averages the scalar product of all pairs of velocity fluctuation vectors separated by a distance $R=|{\bf R}|$, 
\begin{equation}
C_{{\bf u}{\bf u}}(R)\equiv\left<\frac{\sum_{j}\delta{\bf u}({\bf r}_{j})\cdot\delta{\bf u}({\bf r}_{j}+{\bf R})}{\sum_{j}\delta{\bf u}({\bf r}_{j})\cdot\delta{\bf u}({\bf r}_{j})}\right>
\end{equation}
The angle brackets signify an average over all $xy$ directions, and the sums are taken over all positions ${\bf r}_{j}$ \cite{tommy,swirl}.  For small $R$, $C_{{\bf u}{\bf u}}(R)$ decays precipitously from one, as shown by the blue circles in Fig.~3; this decay is nearly exponential [Fig.~3, inset], with a characteristic length scale of order one pore \cite{2dsims}. Intriguingly, however, we also observe weak oscillations in $C_{{\bf u}{\bf u}}(R)$ at even larger $R$; this indicates the presence of slight, but non-zero, correlations in the flow that persist up to distances spanning several pores. We hypothesize that these oscillations reflect the geometry of the pore space formed by the packing of the beads. To test this idea, we compare the shape of $C_{{\bf u}{\bf u}}(R)$ with that of the pore-space pair correlation function, $f(R)$, of a random packing of beads similar to that comprising our porous medium; this function describes the probability of finding a point of the pore space at a distance $R$ away from another point in the pore space. Similar to $C_{{\bf u}{\bf u}}(R)$, $f(R)$ also shows oscillations \cite{audoly}; these reflect the local packing geometry of the spherical beads \cite{nelson}. Moreover, the peaks in $C_{{\bf u}{\bf u}}(R)$ occur at $R\approx$ 1, 2, 2.8, and 3.7 bead diameters, as indicated in the inset to Fig.~3, in excellent coincidence with those observed in $f(R)$ \cite{supp}. This close agreement confirms that the correlations in the flow are determined by the geometry of the pore space.

 \begin{figure}
\begin{center}
\includegraphics[width=3.0in]{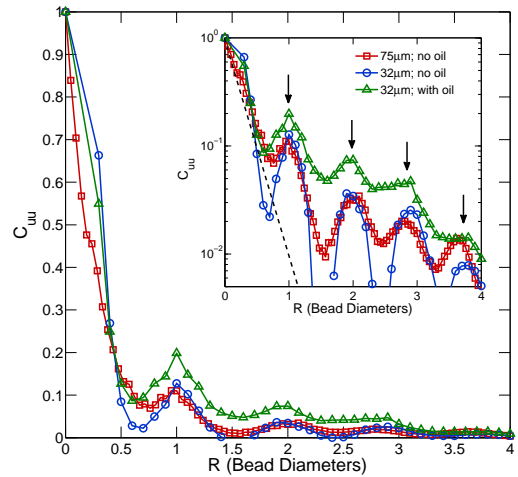}
\caption{Spatial correlation function of velocity fluctuation vectors, $C_{{\bf u}{\bf u}}(R)$, decays with distance $R$. Blue circles and red squares are for single-phase flow through media comprised of $32~\upmu$m and $75~\upmu$m radius beads, respectively, while green triangles are for flow through the medium comprised of $32~\upmu$m radius beads, with trapped residual oil. Inset shows the same data, plotted with semilogarithmic axes; dashed line indicates an exponential decay with characteristic lengthscale $\sim$ 1 pore. Arrows indicate positions of peaks in $C_{{\bf u}{\bf u}}$.}  
\end{center}
\end{figure}

This disordered geometry forces each fluid element to follow a tortuous path through the medium, traveling a total distance larger than $L$. Averaging the distances traveled by all the fluid elements yields an effective distance traveled $\sqrt{\alpha} L$, where $\alpha>1$, often referred to as the hydrodynamic tortuosity, provides an important and commonly-used measure of the variability of the flow. Acoustic \cite{Johnson}, electrical \cite{Johnson}, pressure \cite{Charlaix1}, NMR \cite{Walsworth, Gladden, Davies}, and dispersion \cite{Charlaix2} measurements, as well as a theoretical calculation \cite{Pabitra}, yield $\alpha\approx2$ for a porous medium similar to ours. Within the picture presented here, the distance traveled by each fluid element is approximately $L/\cos^{2}\theta$ \cite{angles}; we thus use our measured velocity orientations [blue circles in Fig.~2(b)] to directly calculate the tortuosity. We find $\alpha=1.80$, in good agreement with the previously obtained values. This provides additional confirmation of the validity of our picture. 

To test the generality of our results, we perform similar measurements on another 3D porous medium with beads of larger radii, $75\pm4~\upmu$m. The average interstitial velocity of the imposed flow is $q/\phi=34~\upmu$m/s. Similar to the case of the smaller beads, we observe broad, exponentially-decaying velocity pdfs [red squares in Fig.~2]; moreover, the pdfs for both porous media collapse when the velocities are rescaled by $q/\phi$. We again quantify the spatial correlations in the flow using the function $C_{{\bf u}{\bf u}}(R)$. As in the case of the smaller beads, $C_{{\bf u}{\bf u}}(R)$ decays exponentially for $R<1$ bead diameter, and also exhibits slight oscillations at even larger $R$, as shown by the red squares in Fig.~3. The close agreement between the measurements on both porous media confirms that our results are more general.

Many important situations, such as oil recovery, groundwater contamination, and geological CO$_{2}$ storage, involve flow around discrete ganglia of a second, immiscible, fluid trapped within the pore space \cite{bear}. This trapping dramatically alters the continuum transport, presumably due to modifications in the pore scale flow \cite{amber,okamoto,degennes}. However, investigations of this behavior are woefully lacking; scattering of light from the ganglia surfaces typically precludes direct visualization of the tracer-laden fluid flow around them. We overcome this challenge by formulating a second non-wetting fluid composed of a mixture of hydrocarbon oils; this composition is carefully chosen to match its refractive index to that of the wetting tracer-laden fluid and the glass beads, thereby enabling full visualization of the tracer-laden fluid flow \cite{amber,supp}. To trap residual ganglia of the oil, we flow it for $30$ min at a rate of $10~$mL/hr through the porous medium comprised of the smaller beads; we then reflow the tracer-laden fluid at a rate of $0.1~$mL/hr, corresponding to a capillary number Ca$~\equiv\mu q/\gamma\approx10^{-5}$, where $\gamma\approx13~$mN/m is the interfacial tension between the two fluids. This protocol leads to the formation of discrete ganglia that remain trapped within the pore space, as indicated in Fig.~1(c) \cite{morrow}. The tracer-laden fluid continues to flow around these ganglia; we directly visualize this steady-state flow using confocal microscopy, re-acquiring movies of optical sections at the same positions as those obtained prior to oil trapping. 

Similar to the previous case without residual trapping, the flow is highly variable, as illustrated by the map of velocity magnitudes shown in Fig.~1(c). Because the ganglia occlude some of the pore space, the characteristic speed of the tracer-laden fluid is larger, $\approx 1.2q/\phi$ [green triangles in Fig.~2(a)]; moreover, because the tracer-laden fluid must flow around the ganglia, more velocities are oriented transverse to the flow direction [green triangles in Fig.~2(b)]. As in the case without residual trapping, we observe broad, exponentially-decaying pdfs for the velocity components [green triangles in Fig.~2(c-d)]; however, these pdfs are significantly broader, indicating that residual trapping introduces additional variability to flow within a 3D porous medium. We again use the measured velocity orientations [green triangles in Fig.~2(b)] to directly calculate the tortuosity, $\alpha$. Consistent with previous indirect measurements \cite{burdine}, we find $\alpha=2.24$, higher than the tortuosity measured in the previous case of single-phase flow; this further reflects the additional flow variability introduced by residual trapping.

We quantify the spatial correlations in this flow using the function $C_{{\bf u}{\bf u}}(R)$. Interestingly, as in the previous case without residual trapping, $C_{{\bf u}{\bf u}}(R)$ decays exponentially for $R<1$ bead diameter, also exhibiting slight oscillations for even larger $R$ at the same positions, as shown by the green triangles in Fig.~3. This indicates that the flow remains correlated, even when a second immiscible fluid is trapped within the medium; moreover, the structure of these correlations is again determined by the geometry of the pore space. 

Our measurements quantify the strong velocity variations in single- and multi-phase flow within a 3D porous medium. We find that the velocity magnitudes and the velocity components both along and transverse to the imposed flow direction are exponentially distributed. Moreover, we present direct evidence that the flow is correlated at the pore scale, and that the structure of these correlations is determined by the geometry of the medium. The pore space is highly disordered and complex; nevertheless, our results suggest that flow through it is not completely random.

This work was supported by the NSF (DMR-1006546), the Harvard MRSEC (DMR-0820484), and the Advanced Energy Consortium (http://www.beg.utexas.edu/aec/), whose member companies include BP America Inc., BG Group, Petrobras, Schlumberger, Shell, and Total. SSD acknowledges funding from ConocoPhillips. HC acknowledges funding from the Harvard College Research Program. It is a pleasure to acknowledge the anonymous reviewers for useful feedback on the manuscript; T. E. Angelini, M. P. Brenner, D. L. Johnson, L. Mahadevan, and J. R. Rice for stimulating discussions; A. Pegoraro for assistance with the correlation analysis; S. Ramakrishnan for experimental assistance; W. Thielicke and E. J. Stamhuis for developing and releasing the PIVlab package; and W. Thielicke for assistance with PIVlab.


\begin{thebibliography}{55}
\bibitem[{tai()}]{tail}
\bibinfo{note}{We note that the pdf for $u_{L}$ shows a small tail of negative
  velocities, possibly due to recirculation, as has been observed in previous NMR measurements \cite{Kutsovsky}.}
\bibitem[{ang()}]{angles}
\bibinfo{note}{Because the porous medium is isotropic, we assume that the pdf of $\theta$ also provides the pdf of the solid angle in the $xz$ plane.}
\end{thebibliography}
\end{document}


\title{Supporting Information: \\Spatial Fluctuations Of Fluid Velocities \\In Flow Through A Three-Dimensional Porous Medium}

    \author{Sujit S. Datta}
    \affiliation{Department of Physics, Harvard University, Cambridge MA 02138, USA}           

    \author{H. Chiang}
            \affiliation{Department of Physics, Harvard University, Cambridge MA 02138, USA}

    \author{T. S. Ramakrishnan}
    \affiliation{Schlumberger-Doll Research, Cambridge, MA 02138, USA}

    \author{David A. Weitz}
    \affiliation{Department of Physics, Harvard University, Cambridge MA 02138, USA}
\email{weitz@seas.harvard.edu}
\pacs{47.56.+r,47.61.-k,47.55.-t}
\maketitle

We prepare rigid 3D porous media by lightly sintering dense, disordered packings of hydrophilic glass beads, with radii $32\pm2~\upmu$m or $75\pm4~\upmu$m, in thin-walled square quartz capillaries of cross-sectional area $A=9$ mm$^{2}$ or $4$ mm$^{2}$, respectively. The packings have length $L\approx8$ mm and porosity $\phi\approx0.41$, as measured using confocal microscopy; this corresponds to a random loose packing of frictional particles. Scattering of light from the surfaces of the beads typically precludes direct observation of the flow within the medium. We overcome this limitation by formulating a mixture of $82~\mbox{wt}$\% glycerol, $12~\mbox{wt}$\% dimethyl sulfoxide, and $6~\mbox{wt}$\% water, laden with 0.01 vol\% of 1 $\upmu$m diameter fluorescent latex microparticles; this composition matches the fluid refractive index with that of the glass beads, enabling full visualization of the flow through the pore space \cite{amber}. Without particles, the density of the tracer fluid is $\rho_{w}\approx1.23$ g/cm$^{3}$, while the density of the particles is $\rho_{p}\approx1.01$ g/cm$^{3}$. The dynamic viscosity of the tracer fluid, without particles, is $\mu\approx60$ mPa$~$s.

Prior to each experiment, the porous medium is evacuated under vacuum and saturated with CO$_{2}$ gas; this gas is soluble in the tracer fluid and prevents the formation of any trapped bubbles. We then saturate the pore space with the tracer fluid, imposing a constant volumetric flow rate $Q=0.2$ mL/hr; the ``superficial" velocities are thus $q/\phi\equiv(Q/A)/\phi=15~\upmu$m/s and $34~\upmu$m/s for the media comprised of beads with radius $32~\upmu$m and $75~\upmu$m, respectively. The typical Reynolds number of the pore-scale flow is thus Re$~\equiv\rho_{w} qa/\mu\approx10^{-4}-10^{-3}$.

To directly visualize the steady-state pore-scale flow, we use a confocal microscope to acquire a movie of 100 optical slices in the $xy$ plane, collecting 15 slices/s, at a fixed $z$ position several bead diameters deep within the porous medium. Each slice is 11 $\upmu$m thick along the $z$ axis and spans a lateral area of 912 $\upmu$m $\times$ 912 $\upmu$m in the $xy$ plane. To visualize the flow at the scale of the entire medium, we acquire additional movies, at the same $z$ position, but at multiple locations in the $xy$ plane spanning the entire width and length of the medium. We restrict our analysis to an area several beads away from each edge of the medium to minimize boundary effects. To reduce image noise, we threshold and apply a median filter to each optical slice. We have verified that the results presented do not significantly change depending on how the images are filtered. We have also verified that the results do not change with depth ($z$ position), for the imaging parameters used. 

The ratio of viscous forces to gravitational forces on the tracer particles is given by $9\mu (q/\phi)/2(\rho_{w}-\rho_{p})ga_{p}^{2}\approx10^{4}-10^{5}$, where $g$ is gravitational acceleration; moreover, the time required for a particle to settle a vertical distance of $11\mu$m, equal to the thickness of the optical slice, is $<10^{4}\times11~\upmu\mbox{m}/(q/\phi)\sim10$ min, approximately 100 times larger than the time required to acquire one movie. The time for a tracer particle to diffuse its own size is approximately $\pi\mu a_{p}^{3}/k_{B}T\approx6$ s, where $k_{B}$ is Boltzmann's constant and $T\approx300$K is temperature; this is almost 100 times larger than the time between two successive frames of the movies. We thus conclude that the tracer particles are faithfully advected with the flow. We note that measurements in two dimensional micromodels suggest that the tracer particles may not fully sample the smallest fluid velocities near the bead surfaces \cite{bechinger}. However, we do not expect this to significantly affect the measured distribution of the larger fluid velocities, which are the primary focus of our paper. Moreover, we do not observe significant differences between the data for porous media comprised of $32~\upmu$m and $75~\upmu$m beads. The tracer Peclet number, quantifying the importance of advection relative to diffusion in determining the particle motion, is $(q/\phi)/(k_{B}T/6\pi a_{p})>10^{3}$; thus, our experiments only probe the advection motion of the fluid, in contrast to other measurements that measure the combined effects of advection and diffusion \cite{scheven}. 

We use the PIVLab package for MATLAB to characterize the flow field using particle image velocimetry, dividing each optical slice into 16129 interrogation windows, and calculating the displacement of tracer particles in each window by cross-correlating successive slices of each movie. By combining the displacement field thus obtained for all the positions imaged, and dividing the displacements by the fixed time difference between slices, we generate a map of the fluid velocity over the entire porous medium. Thus, this protocol enables us to directly visualize the flow field, both at the scale of the individual pores and the overall medium. For calibration, we use this protocol to visualize and confirm the parabolic Poiseuille flow profile within a square capillary of cross-sectional width $550~\upmu$m. Moreover, we confirm that PIV performed using interrogation windows 0.75 and 0.5 times the size of those used here yield similar results.

 \begin{figure}
\begin{center}
\includegraphics[width=3.5in]{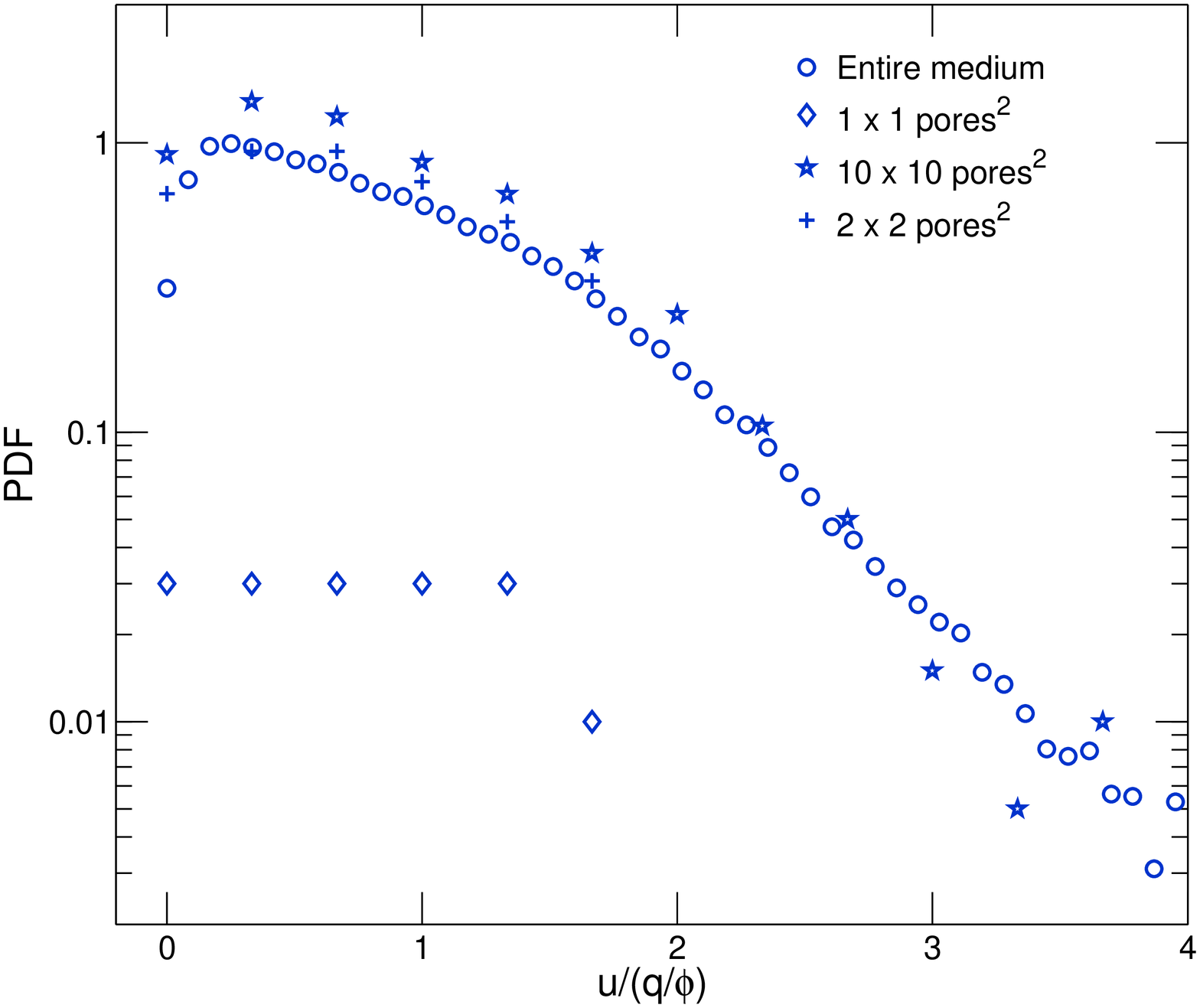}
\caption{Probability density functions of normalized 2D velocity magnitudes, $u/(q/\phi)$, for single-phase flow without residual oil trapping through a medium comprised of beads with average radius $32~\upmu$m, for square windows approximately one (diamonds), two (crosses), and ten (stars) pores across; circles show data for the entire porous medium. Data are vertically offset for clarity. } 
\end{center}
\end{figure}

 \begin{figure}
\begin{center}
\includegraphics[width=3.5in]{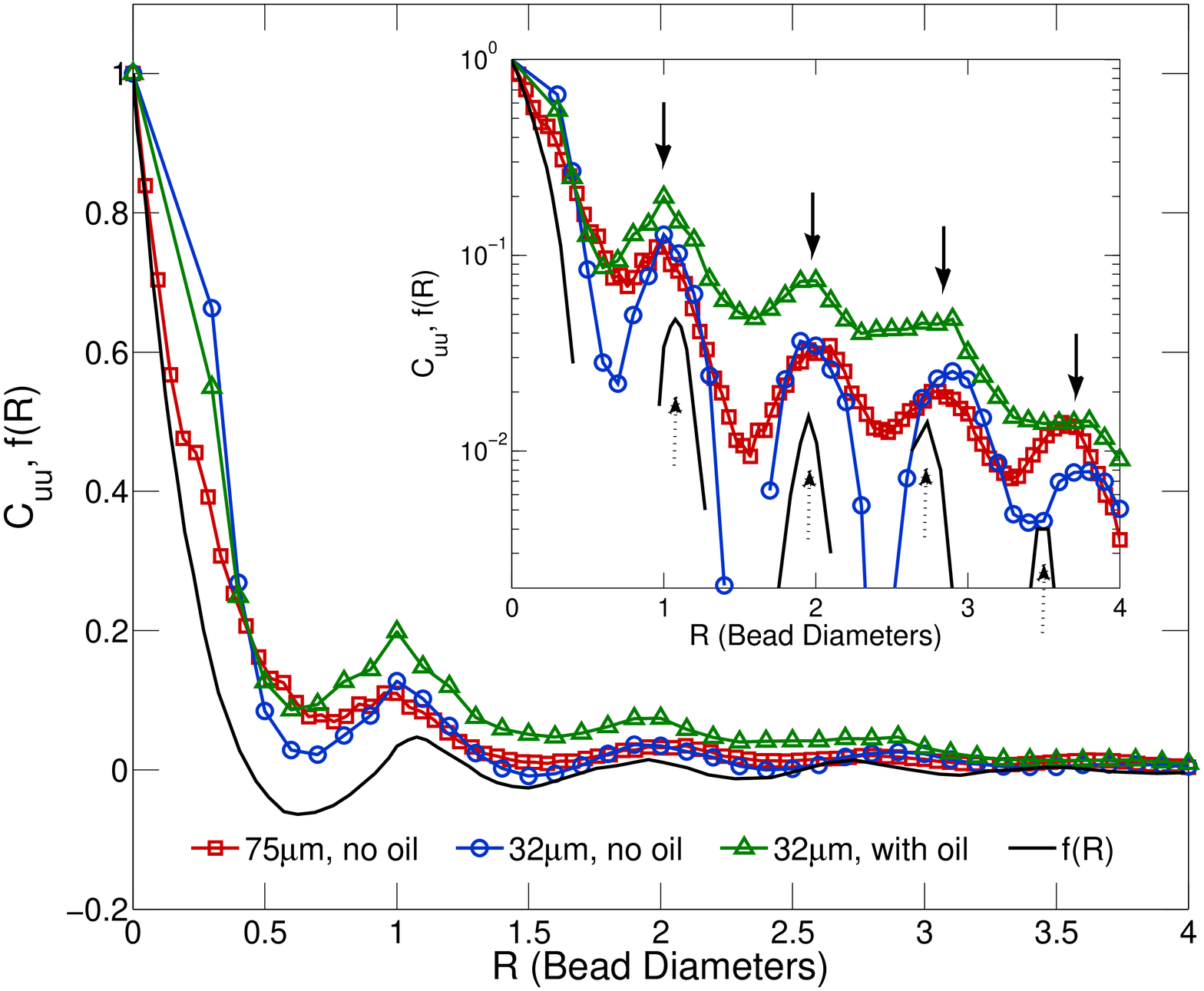}
\caption{Variation of spatial correlation function of velocity fluctuation vectors, $C_{{\bf u}{\bf u}}(R)$, or pore-space pair correlation function, $f(R)$, with distance $R$. Blue circles and red squares are for single-phase flow through media comprised of $32~\upmu$m and $75~\upmu$m radius beads, respectively, while green triangles are for flow through the medium comprised of $32~\upmu$m radius beads, with trapped residual oil. Data for $f(R)$ are taken from \cite{audoly}. Inset shows the same data, plotted with semilogarithmic axes. Solid arrows indicate positions of peaks in $C_{{\bf u}{\bf u}}$ at $R\approx$ 1, 2, 2.8, and 3.7 bead diameters, while dashed arrows indicate positions of peaks in $f(R)$ at $R\approx$ 1.1, 2, 2.7, and 3.5 bead diameters. Both $C_{{\bf u}{\bf u}}$ and $f(R)$ show similar oscillatory behavior, with peaks and troughs at similar locations. } 
\end{center}
\end{figure}